\documentclass[notitlepage,12pt,nofootinbib,tightenlines,prb]{revtex4-2} %
\usepackage{setspace}
\usepackage{amsmath}
\usepackage{amsfonts}
\usepackage{graphicx}
\usepackage{cancel}
\usepackage[hidelinks]{hyperref}
\usepackage{todonotes}

\newcommand{\order}[1]{\mathcal{O}\left(#1\right)}

\newcommand{\non}{\nonumber\\}

\begin{document}
%\begin{singlespace}
	\title{The Yukawa Interaction in Ordinary Quantum Mechanics}
%	\date{\today}
	\author{Daniel Schubring}
\email{schub071@umn.edu}
\affiliation{Department of Physics, University of Minnesota, Minneapolis, MN 55455, USA}
	\begin{abstract}
		The Yukawa interaction is considered in 0+1 dimensions as a pedagogical example to illustrate quantum field theory methods. From the quantum mechanical point of view the system is trivially exactly solvable, but this can be difficult to see from the path integral perspective. It is shown how despite initial appearances the perturbation series in terms of Feynman diagrams is consistent with the quantum mechanical picture for finite temperature.
	\end{abstract}
	
	\maketitle
	\tableofcontents
\begin{spacing}{1.25}
	\section{Introduction}
	Quantum field theory can be a challenging subject to learn and teach, but one approach towards quantum field theory pedagogy is to first focus on fields in one spacetime dimension ($d=1$). This has the advantage that it may be connected to ordinary quantum mechanical systems which may be solved in terms of methods which may already be familiar to a student, such as the use of wave functions. Certain features of quantum field theory calculations, such as the appearance of divergences and use of Grassmann numbers, may meet with some initial skepticism. So it may aid understanding to first introduce these concepts in a system which may be understood from an alternate, more familiar perspective. This also has the benefit of allowing a student to practice the new technical details of quantum field theory, such as the combinatorics of Feynman diagrams, on a problem which they may already be confident in solving a different way.
	
	Unfortunately this article can not take such a broad scope as to teach the reader these basic technical details for the first time. Rather it is directed either towards teachers of quantum field theory who are looking for an enlightening example for their course, or for students who may have already read a textbook on quantum field theory (e.g. \cite{Peskin:1995ev,Montvay:1994cy,Nakahara:2003nw}) and have already been exposed to Feynman diagrams and path integrals, but are looking to solidify their understanding.
	
	 A similar approach towards introducing quantum field theory methods by considering fields in $d=1$ has already been considered by Boozer \cite{Boozer} who discussed both the $\phi^4$ interaction, which may be understood as an anharmonic oscillator in ordinary quantum mechanics, as well the Yukawa interaction, a system which involves fermions and which will be introduced more thoroughly below. A fundamental difference from Boozer's approach is that we shall consider this from the point of view of path integrals in Euclidean time rather than the point of view of canonical quantization. 

	In particular, the treatment of fermions in terms of path integrals over Grassmann numbers may be something that can seem quite disconnected from previously learned methods in physics, and it may be beneficial to introduce it in terms of a system that may be easily solved in ordinary quantum mechanics. Of course in one spacetime dimension there is no non-trivial Lorentz group so fermion fields will lack a natural spinor index, but otherwise their propagator takes a very similar form to that in higher dimensional field theories. An interacting theory such as we will present here will involve expansions in terms of Feynman diagrams which will take essentially the exact same form as those in genuine quantum field theories in higher dimension. But from the perspective of canonical quantization, a single fermion field in $d=1$ is just associated to a ladder operator in a two-level system. So it is interesting to see how this simple two-level picture could possibly be compatible with all the complexity of fermion propagators in an asymptotic series of Feynman diagrams. Seeing how this works out explicitly is to a large extent the main goal of the present article.
	
	Part of the motivation for considering $d=1$ fermions in more detail here comes from a related study of constrained quantum mechanical systems such as the quantum rotor from a quantum field theory perspective \cite{1dO(n)}. There it was pointed out that from the perspective of canonical quantization the fermionic corrections to the bosonic correlation functions of the supersymmetric quantum rotor should collectively vanish, even though this does not occur in higher spacetime dimension. So it was interesting to see what special features of fermions in $d=1$ lead to such a cancellation from a purely path integral perspective.
	
	But to better understand fermionic fields in $d=1$, a far simpler model than a supersymmetric non-linear sigma model may be considered. Perhaps the simplest interacting theory involving a scalar field $\phi$ and fermionic field $\bar{\psi},\psi$, is just the \emph{Yukawa interaction} which has the following action integrated over Euclidean time $\tau$,
	\begin{align}
		S = \int d\tau\, \left[\frac{1}{2}\left((\partial_\tau \phi)^2 + m^2 \phi^2\right)+\bar{\psi}\left(\partial_\tau + \mu\right)\psi + \lambda \phi \bar{\psi}\psi\right].\label{Action}
	\end{align}
	This just has the usual quadratic terms for a scalar and fermion field extrapolated down to $d=1$, and an interaction term $\phi\bar{\psi}\psi$ which is the main characteristic of the model. The fields $\phi$ and $\psi$ have `bare masses' $m$ and $\mu$ respectively. The derivative term for the fermion field may appear a little unusual but this is just due to the triviality of one-dimensional Clifford algebras and lack of spinor indices. If it is desired to emphasize the connection to higher dimensional theories, a single one component gamma `matrix' $\gamma^\tau=i$ may be introduced in terms of which fermionic derivative term may also be expressed as $-i\bar{\psi}\gamma^\tau \partial_\tau \psi$. Note that we are setting $\hbar=1$ so that the action is dimensionless. Given that $\tau$ has dimensions of length, this implies that the the scalar field $\phi$ has dimensions of the square root of length, $\psi$ is dimensionless, and the coefficient $\lambda$  of the interaction term must have dimensions of length to the $-3/2$ power. 
	
	Now let us consider this system from the point of view of ordinary quantum mechanics. The connection between operator formalism and the Euclidean path integral formalism for $d=1$ is reviewed for instance in Chapter 1 of \cite{Nakahara:2003nw}.
	The action $S$ is associated to a  a Hamiltonian $H$ expressed in terms of position operator $q$, momentum operator $p$, and fermionic creation and annihilation operators $c^\dagger, c$,
	\begin{align}
		H= \frac{1}{2}\left(p^2+m^2 q^2\right)+\mu c^\dagger c + \lambda q c^\dagger c.\label{Hamiltonian}
	\end{align}
Actually the ordering of the operators in the interaction term is not specified fully by the equal time $\phi\bar{\psi}\psi$ term in the path integral. We shall return to this issue later, but for now we are simply taking this ordering as given and will choose our regularization in the path integral approach in order to ensure this holds.
	
	From this Hamiltonian we can see that the Yukawa interaction is actually extremely simple in $d=1$. The Hamiltonian commutes with the number operator $N\equiv c^\dagger c$, so we may consider the $N=0$ and $N=1$ eigenspaces separately. For states with $N=0$, which may be called the \emph{bosonic sector}, the Hamiltonian reduces to an ordinary harmonic oscillator Hamiltonian with angular frequency $m$
	\begin{align}
		H_B = \frac{1}{2}\left(p^2+m^2 q^2\right),
	\end{align}
which has spectrum $E_{n,B}=\left(n+\frac{1}{2}\right)m.$ For the \emph{fermionic sector} with $N=1$, the Hamiltonian $H_F$ just reduces to a shifted harmonic oscillator by completing the square,
\begin{align}
H_F = \frac{1}{2}\left(p^2+m^2 \left(q+\frac{\lambda}{m^2}\right)^2\right)+\mu-\frac{\lambda^2}{2m^2}.
\end{align}
Here the constant shift in the Hamiltonian may be considered the energy $\mu_\lambda$ of the fermion itself
\begin{align}
	\mu_\lambda \equiv \mu-\frac{\lambda^2}{2m^2}.\label{mlambda}
\end{align}

The spectrum in the fermion sector is just $E_{n,F}=E_{n,B}+\mu_\lambda$, so as far as the spectrum is concerned the Yukawa interaction only serves to correct the energy of the fermionic excitation from $\mu$ to $\mu_\lambda$. The interaction also leads to a non-zero expectation value for $q$, since the minimum of the harmonic oscillator potential is shifted,
\begin{align}
	\langle q\rangle_F = -\frac{\lambda}{m^2}.\label{qF}
\end{align}

Note that if $\lambda$ is large enough so that $\mu_\lambda$ becomes negative, then the true ground state is in the fermionic sector. From the quantum field theory perspective this sounds a bit more impressive. As $\lambda$ increases the system goes into a phase where the $\bar{\psi}\psi$ operator `condenses' and the bosonic field $\phi$ picks up a vacuum expectation value. But from the quantum mechanical perspective this is all very straightforward.

It is now our goal to see how the simple quantum mechanical picture described above is consistent with the correlation functions calculated perturbatively from the path integral. The demonstration of this will be organized as follows. First in Sec. \ref{section correlation functions} we will very briefly review the non-perturbative connection between Euclidean correlation functions and the spectrum of the underlying quantum mechanics. In Sec. \ref{section Zero temperature} we move on to calculating these correlation functions in the vacuum at zero temperature. First the technical details of regularization and its relation to operator ordering must be considered, but then the expressions for $\mu_\lambda$ and $\langle q\rangle_F$ above will be rederived through a Feynman diagram calculation.

One of peculiarities of fermions in $d=1$ from the path integral perspective will be shown to be the vanishing of all loops involving more than one fermion propagator. This may be rather puzzling at first because the perturbative expansion will involves all sorts of diagrams involving fermion loops, and it seems that these must be playing some role in the theory.

This issue will be clarified in Sec. \ref{section: finite temp} by considering correlation functions at non-zero temperature where the fermion loops no longer vanish. The fermion pair correlation functions investigated here are familiar from many-body quantum mechanics (see e.g. \cite{pairCorrelation}) and it is shown how a delicate cancelation takes place in order for the fermion field bilinear $\bar{\psi}\psi$ to be consistent with its quantum mechanical interpretation as a number operator in a simple two-level system.  Finally an alternative summary of the results of this paper is given in Sec. \ref{section conclusion}, and readers may freely skip ahead.

\section{Euclidean correlation functions}\label{section correlation functions}
The path integral approach gives a natural way to calculate correlation functions of operators which are ordered in Euclidean time. This section will review how information about the quantum mechanical spectrum may be extracted from the two-point correlation functions. This is essentially a simpler version of the same idea that underlies the well-known K\"{a}ll\'{e}n-Lehmann spectral representation in standard quantum field theory, which is treated for instance in Sec. 7.1 of \cite{Peskin:1995ev}. This material is also covered in Boozer's article \cite{Boozer}, and the reader may find the related discussion of the quantum rotor in the first three sections of \cite{1dO(n)} helpful as it spends more time discussing the paradigm shift between quantum mechanics and quantum field theory and begins with zero-dimensional path integrals as a warm up case.

%\footnote{The index $n$ is here taken to include information about both the harmonic oscillator energy level and the eigenvalue of the number operator $c^\dagger c$ }

One advantage of considering Euclidean time $\tau$ rather than ordinary Minkowski time $t=-i\tau$ is due to the formal similarity between the Euclidean time translation operator $e^{-\tau H}$ and the Boltzmann factor involved in thermal expectation values. A thermal expectation value of an operator $\hat{O}$ is given by a trace over eigenstates $|n\rangle$ of the Hamiltonian.
\begin{align}
	\langle \hat{O}\rangle_\beta \equiv \frac{1}{Z(\beta)} \sum_{n} \langle n|\,e^{-\beta H}\hat{O}\,| n\rangle,\qquad Z(\beta)\equiv\sum_{n} \langle n|\,e^{-\beta H}\,| n\rangle.\label{thermal expectation}
\end{align}
In the zero temperature $\beta\rightarrow \infty$ limit this just becomes a vacuum expectation value, and that will be our first focus.

The Euclidean time ordered correlation functions in the bosonic ($N=0$) ground state $|0_B\rangle$ will be enough to tell us the energies of a single bosonic or fermionic excitation. To be clear, the Euclidean time dependence of an operator is given by $\hat{O}(\tau)=e^{H\tau}\hat{O}e^{-H\tau}$. Inserting an identity operator in the form $I=\sum_n |n\rangle\langle n|$ next to a Euclidean field allows us to replace the time translation operator by its eigenvalue acting on the state $|n\rangle$. Using this trick the time-ordered two-point correlation function for $q$ evaluates to
\begin{align}
 \langle 0 |\, \mathcal{T}q(\tau_f)q(\tau_i)\,|0\rangle &= \sum_n \left|\langle n| q|0\rangle\right|^2
 \,e^{-\left(E_n-E_0\right)|\tau_f-\tau_i|}\non&=\int\frac{dp}{2\pi}e^{-i(\tau_f-\tau_i)p}\,\sum_n \frac{2(E_n-E_0)\left|\langle n| q|0\rangle\right|^2}{p^2+\left(E_n-E_0\right)^2}.\label{propagator phi}
\end{align}
The Fourier transformed form in the second line may be verified by direct contour integration in the complex $p$ plane. If $\tau_f-\tau_i > 0$ then we must close the integration contour in the negative imaginary half plane which encloses the pole $p=-i(E_n-E_0)$, and vice versa if $\tau_f-\tau_i < 0$. This correctly produces the absolute value in the exponential arising from the Euclidean time ordering.

So from this expression it is seen that the poles of the two-point function in momentum space give us information about the spectrum $E_n-E_0$. The simple system considered here \eqref{Hamiltonian} is essentially a harmonic oscillator and $q|0_B\rangle$ is exactly the first excited state up to normalization. So all the amplitudes $\left|\langle n| q|0_B\rangle\right|^2$ will vanish except for the level $n=1$. The difference $E_1-E_0$ is just the harmonic oscillator frequency  $m$, and this is uncorrected even in the presence of non-zero $\lambda$.

The fermionic correlation function in the bosonic vacuum follows similarly from the same trick of inserting the identity operator expressed in terms of energy eigenstates 
\begin{align}
 \langle 0_B |\, \mathcal{T}c(\tau_f)c^\dagger(\tau_i)\,|0_B\rangle = \sum_i \left|\langle n| c^\dagger|0_B\rangle\right|^2
\,e^{-\left(E_i-E_0\right)(\tau_f-\tau_i)}\theta(\tau_f-\tau_i).
\end{align}
Here $\theta(\tau_f-\tau_i)$ is a Heaviside function which vanishes if $\tau_f<\tau_i$ since due to the time ordering in that case the annihilation operator $c(\tau_f)$ would appear to the right acting on the vacuum $|0_B\rangle$. As before it may be verified by direct contour integration that this is equal to
\begin{align}
 \langle 0_B |\, \mathcal{T}c(\tau_f)c^\dagger(\tau_i)\,|0_B\rangle = \int\frac{dp}{2\pi}e^{-i(\tau_f-\tau_i)p}\,\sum_n \frac{\left|\langle n| c^\dagger|0_B\rangle\right|^2}{-ip+\left(E_n-E_0\right)}.\label{propagator psi}
\end{align}

Once again the poles give information on the spectrum, but in this case the infinite sum over $n$ will remain since the harmonic oscillator eigenstates in the fermionic sector have shifted wavefunctions. As far as the $q$ and $p$ operators are concerned, the state $c^\dagger |0_B\rangle$ is a Gaussian centered at $q=0$ just like $|0_B\rangle$ itself, so it has non-zero overlap with the excited harmonic oscillator states which are centered at $q=\langle q\rangle_F$ defined in \eqref{qF}. But to lowest order in $\lambda$, $c^\dagger|0_B\rangle$ only has overlap with the fermionic ground state $|0_F\rangle$, and that has an energy difference $E_{0,F}-E_{0,B}$ which is equal to $\mu_\lambda$ in \eqref{mlambda}.

We wish to compare these correlation functions to those found via the path integral approach using action \eqref{Action}. A good introduction to Euclidean path integrals including the finite $\beta$ case that will be discussed in Sec. \ref{section: finite temp} is found for instance in the introduction of \cite{Montvay:1994cy}. In momentum space, the partition function may be written
\begin{gather}
	Z(\beta\rightarrow \infty)=\int\mathcal{D}{\phi}\mathcal{D}\bar{\psi}\mathcal{D}{\psi}\,e^{-S_0-\lambda S_I},\non
	S_0=\int \frac{dp}{2\pi}\left[\frac{1}{2}\phi(-p)\left(p^2+m^2\right)\phi(p)+\bar{\psi}(-p)\left(-ip+\mu\right)\psi(p)\right],\non S_I=\int \frac{dp}{2\pi}\frac{dk}{2\pi}\,\phi(-p)\bar{\psi}(p-k)\psi(k).\label{action momentum space}
\end{gather}
The propagators of $\phi$ and $\psi$ may be read directly from the quadratic action $S_0$ to be $(p^2+m^2)^{-1}$ and $(-ip+\mu)^{-1}$ respectively. Since the harmonic oscillator amplitude $|\langle 1| q |0\rangle|^2$ may be easily calculated as $(2m)^{-1}$ using either wave function or operator methods it is seen that these propagators are compatible with the exact expressions \eqref{propagator phi} and \eqref{propagator psi} in the $\lambda=0$ limit.

In the next section we will show explicitly via path integral methods that indeed the bosonic propagator is uncorrected by the Yukawa interaction $S_I$, and the fermionic mass $\mu$ is corrected to $\mu_\lambda$ in agreement with the quantum mechanical picture. The situation for a finite temperature $\beta$ will be a bit more problematic and will be the focus of Sec. \ref{section: finite temp}.

\section{Path integral approach} \label{section Zero temperature}
%\footnote{In momentum space this integral will also have an overall delta function which is not shown but will be important for the finite temperature case later.} 
\subsection{Expectation value of $\phi$ and ordering ambiguities}\label{section Tadpole}
\begin{figure}
	\centering
	\includegraphics[width=0.8\textwidth]{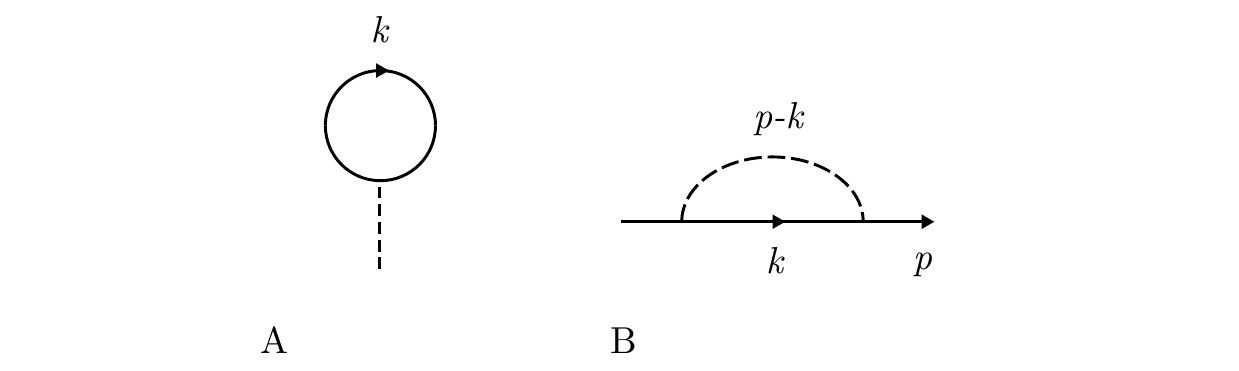}
	\caption{The lowest order corrections $\langle \phi \rangle^{(1)}$ and $\langle \psi(p)\bar{\psi}(-p)\rangle^{(2)}$ are shown in A and B respectively. Dashed lines denote $\phi$ propagators and solid lines denote $\psi$ propagators.}\label{Fig1}
\end{figure}

Let us first consider the expectation value of $\phi$ in the path integral approach. From \eqref{qF} we expect that this will vanish in the bosonic vacuum and equal $-\lambda/m^2$ in the fermionic vacuum. Denoting the order in powers of $\lambda$ by a superscript, the first order correction $\langle\phi\rangle^{(1)}$ will involve a single interaction vertex $-\lambda S_I$ with both fermion fields contracted with each other. This correction is represented as the tadpole diagram in part A of Fig. \ref{Fig1}. Recall that fermion loops such as this also lead to an overall minus sign due to reordering the Grassmann number fields. The correction leads to a logarithmically divergent integral
\begin{align*}
	\langle \phi \rangle^{(1)} = +\frac{\lambda}{m^2}\int \frac{dk}{2\pi}\frac{1}{-ik+\mu}.
\end{align*}

The logarithmic divergence has a clear interpretation. It is a manifestation of the ordering ambiguity for the equal time product of fields $\bar{\psi}(\tau)\psi(\tau)$ in the interaction vertex. The operators $c^\dagger, c$ have non-zero anticommutator, so it is ambiguous whether correlation functions involving $\bar{\psi}\psi$ should mean $c^\dagger c, -cc^\dagger$ or some linear combination of the two when considered as quantum mechanical expectation values. If the ordering ambiguity is fixed by considering the product in the interaction vertex to be $\bar{\psi}(\tau+\epsilon)\psi(\tau)$, for some small positive $\epsilon$, then in momentum space this leads to an extra exponential factor at the interaction vertex \eqref{action momentum space}, $\phi(0)\bar{\psi}(-k)\psi(k)e^{+ik\epsilon}$. So the integral should be regularized to
\begin{align}
	\langle \phi \rangle^{(1)} = \lim_{\epsilon\rightarrow 0^+}\frac{\lambda}{m^2}\int \frac{dk}{2\pi}\frac{e^{+ik\epsilon}}{-ik+\mu}.\label{phi tadpole}
\end{align}
If $\mu$ is taken as positive, so that the ground state is the bosonic vacuum, then the pole in $k$ is in the lower half of the complex plane. But in order for the $e^{+ik\epsilon}$ factor to be convergent, the contour must encircle the upper half plane, where there are no poles. So the expectation value $\langle\phi\rangle_B^{(1)}$ in the bosonic vacuum vanishes.

If $\mu$ is taken to be negative, then the ground state is in the fermionic sector. In terms of the integral, the pole shifts to the upper half plane where it is encircled by the integration countour. The $2\pi i$ from Cauchy's residue theorem and the $(-2\pi i)^{-1}$ from the integral of the propagator itself combine to lead to an overall factor of $-1$. So the expectation value in the fermionic vacuum is $$\langle\phi\rangle_F^{(1)}=-\frac{\lambda}{m^2},$$ which is consistent with the exact result from the quantum mechanical perspective \eqref{qF}.

Finally, note that the time splitting regularization whereby $\bar{\psi}(\tau)\psi(\tau)$ is interpreted as $\lim_{\epsilon\rightarrow 0^+}\bar{\psi}(\tau+\epsilon)\psi(\tau)$ was consistent with our choice of Hamiltonian \eqref{Hamiltonian}, but depending on the system it may not always lead to the desired quantum mechanics. In particular, in considering supersymmetric quantum mechanics from a path integral approach (see e.g. \cite{Cooper:1994eh}) it makes more sense to consider the product $\bar{\psi}\psi$ to represent the commutator $\frac{1}{2}[c^\dagger,c]$. This regularization arises naturally from considering modes of the Dirac operator acting on $\psi(\tau)$ \cite{Gildener:1977hm}\cite{Cooper:1994eh}. It also arises from the simple formal manipulation
$$\int \frac{dk}{2\pi}\frac{1}{-ik+\mu}=\int\frac{dk}{2\pi}\frac{ik+\mu}{k^2+\mu^2}=\frac{1}{2}.$$
Here the odd term $ik$ in the numerator in the second equality is taken to vanish since it is being integrated over a symmetric interval, and the remaining integral is convergent. This seemingly naive form of regularization is actually closely related to dimensional regularization and it was considered in \cite{1dO(n)} for the supersymmetric $O(N)$ model, where a time splitting regularization would not be consistent with the supersymmetry.

\subsection{Fermionic two-point function}\label{section fermion 2 point}

Now we shall consider the correction to the two-point correlation function $S(p)$
\begin{align*}
	S(p)\equiv\int d\tau\, e^{ip\tau}\langle\psi(\tau)\bar{\psi}(0)\rangle,
\end{align*}
	 which will determine the correction to the bare mass $\mu$ and test the first excited states in the fermionic sector. This calculation was earlier considered in \cite{Boozer} from a canonical quantization point of view.
%\footnote{For $\mu<0$ the tadpole in diagram A does not vanish and must also be inserted as a correction to the fermion propagator. This will ensure that the correction $\delta\mu$ has the same sign as for the $\mu>0$ case.}
The lowest order correction involves two interaction vertices and for $\mu>0$ is given by diagram B of Fig. \ref{Fig1} (The $\mu<0$ case involves also a non-vanishing tadpole correction, but is otherwise similar). Diagram B is associated to the integral
\begin{align*}
S(p)^{(2)}&=\frac{\lambda^2}{\left(-ip+\mu\right)^2}\int \frac{dk}{2\pi}\frac{1}{\left(-ik+\mu\right)\left((p-k)^2+m^2\right)}
\end{align*}
This time the integral is convergent without any additional regularization, and the integration contour may be closed in the upper half plane encircling the pole $k=p+im$.
\begin{align}
	S(p)^{(2)}&=\frac{\lambda^2}{\left(-ip+\mu\right)^2}\left[\frac{1}{2m}\frac{1}{-ip+\mu+m}\right].\label{Sp evaluation}
\end{align}
Due to the pole of the self-energy factor, this might at first appear to shift the fermion mass from $\mu$ to $\mu+m$, which does not agree with the exact mass correction $\delta\mu\equiv\mu_\lambda-\mu$. But recall from the non-perturbative form of the correlation function \eqref{propagator psi} that there will be poles associated to each state $|n\rangle$ for which $\langle n|c^\dagger |0_B\rangle\neq 0$. The state $|1_F\rangle$, which may be thought of as involving both a bosonic and a fermionic excitation, has mass $\mu+m$ at lowest order. And since the amplitude $Z_{1,F}\equiv\left|\langle 1_F| c^\dagger|0_B\rangle\right|^2$ is itself of order $\sim\lambda^2$, no corrections to the pole $p=\mu+m$ will be seen in $S(p)$ to this order.

Since the state $c^\dagger|0_B\rangle$ has unit norm, we must have $\sum_n\left|\langle n| c^\dagger|0_B\rangle\right|^2=1$, and at this order this implies that $\left|\langle 0_F| c^\dagger|0_B\rangle\right|^2=1-Z_{1,F}$.
\begin{align}
	S(p)&=\frac{1-Z_{1,F}}{-ip+\mu+\delta\mu}+\frac{Z_{1,F}}{-ip+\mu+m}+\order{\lambda^4}\non
	&=\frac{1}{-ip+\mu}+\frac{-\delta\mu\left(-ip+\mu+m\right)-mZ_{1,F}\left(-ip+\mu\right)}{\left(-ip+\mu\right)^2\left(-ip+\mu+m\right)}+\order{\lambda^4}.
\end{align}

Comparing this expression with the result of the Feynman diagram calculation in \eqref{Sp evaluation}, it is seen that the lack of $p$ dependence in the numerator implies $Z_{1,F}=-\delta \mu/m$. The numerator is then seen to be equal to $-\delta\mu \,m$, which implies
\begin{align}
	\delta\mu=-\frac{\lambda^2}{2m^2},\qquad Z_{1,F}=+\frac{\lambda^2}{2m^3}.\label{psi propagator results}
\end{align}
The mass correction $\delta\mu$ agrees with the exact result \eqref{mlambda}.

The amplitude $Z_{1,F}$ may also be shown to be consistent with the amplitude found from wavefunction methods. Consider the amplitude between the bosonic and fermionic ground states. In both cases the wavefunction is just a Gaussian with some normalization $C$, and from \eqref{qF} the fermionic wavefunction is shifted by $-\lambda/m^2$,
\begin{align*}	\langle 0_F|c^\dagger 0_B\rangle&=C^{2}\int_{-\infty}^{\infty}dq\,e^{-m\frac{\left(q+\frac{\lambda}{m^2}\right)^2}{2}}e^{-m\frac{q^2}{2}}=e^{-\frac{\lambda^2}{4m^3}}.
\end{align*}
And so using $|\langle 0_F|c^\dagger 0_B\rangle|^2=1-Z_{1,F}+\order{\lambda^4}$, we see that this agrees with the result in \eqref{psi propagator results}.

\subsection{Bosonic two-point function}
\begin{figure}
	\centering
	\includegraphics[width=0.8\textwidth]{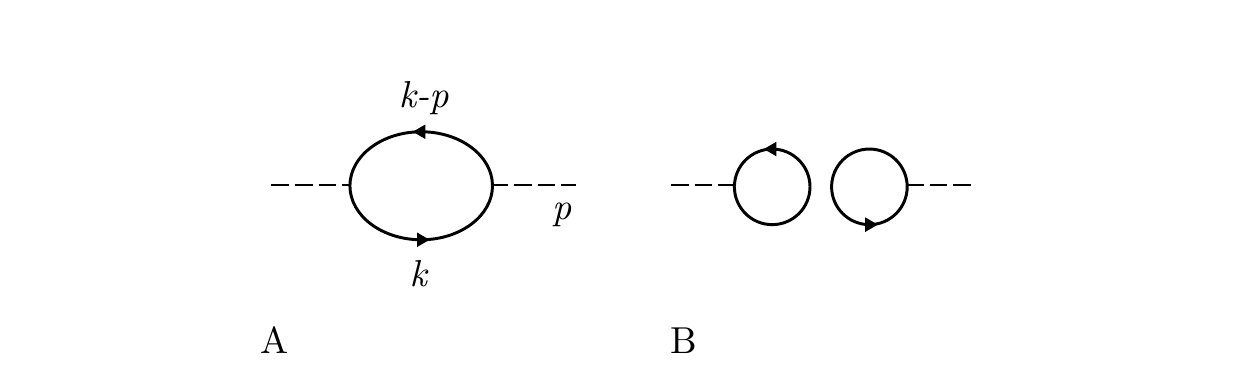}
	\caption{The lowest order corrections $\langle \phi(-p)\phi(p)\rangle^{(2)}$. The connected contribution is given in A, and the disconnected contribution in B involves two copies of diagram A in Fig. \ref{Fig1}.}\label{Fig2}
\end{figure}
Now we will briefly turn to the bosonic two-point function $$D(p)\equiv\int dx\, e^{ip\tau}\langle \phi(\tau)\phi(0)\rangle$$ which will introduce a peculiar feature of fermion loops in $d=1$. The lowest order connected diagram is given in diagram A of Fig. \ref{Fig2},
\begin{align}
	D^{(2)}_\text{connected}=\frac{\lambda^2}{\left(p^2+m^2\right)^2}\int \frac{dk}{2\pi} \left(\frac{1}{k+i\mu}\right)\left(\frac{1}{k-p+i\mu}\right).\label{D connected}
\end{align}
%\footnote{Of course this vanishing may also be seen directly without relying on complex analysis by multiplying the denominators in the integrand by their complex conjugate and integrating over the real line.}
In fact this integral immediately vanishes since both poles in $k$ are in the lower half of the complex plane, and we may close the contour of integration in the upper half plane. This argument will in fact extend to arbitrary loops of fermions such as seen in Fig. \ref{Fig3}, regardless of external momenta. The main point is that \emph{in $d=1$ loops formed entirely of fermions vanish at zero temperature.} As noted above this is not true for the logarithmically divergent tadpole loops involving a single fermion propagator, and the result will depend on the regularization and the sign of $\mu$.

The vanishing of the correction $D^{(2)}_\text{connected}$ is exactly what is needed from the quantum mechanical perspective. In the bosonic vacuum, which holds for $\mu>0$, the Hamiltonian $H_B$ is just an uncorrected harmonic oscillator, so $D$ should receive no corrections in perturbation theory. The disconnected diagram B in Fig \ref{Fig2} will also vanish for $\mu>0$. Conversely when $\mu<0$, $\phi$ picks up a vacuum expectation value, and diagram B will give the proper disconnected correction $\langle\phi\rangle^2$ to the correlation function $\langle\phi(\tau)\phi(0)\rangle$. But even in the fermionic vacuum, the Hamiltonian $H_F$ is just that of a harmonic oscillator with the same frequency $m$, and so there should be no connected corrections to $D$ in perturbation theory.
\section{Considerations at finite temperature}\label{section: finite temp}

The situation above involving the vanishing of all corrections to the bosonic correlation function may seem a little too glib. After all, as the order in the $\lambda$ expansion increases there are more and more potential corrections to the bosonic correlation function which appear and it seems that they must play some role in the theory. To understand the role of these diagrams we need to turn on a finite temperature so that expectation values are taken over a trace of all states and the bosonic fields can `see' the fermionic sector. From the path integral perspective, the finite temperature simply modifies the $\phi$ and $\psi$ fields to be periodic and anti-periodic over the Euclidean time $\beta$, but otherwise the exact same diagrams appear as for the zero-temperature case.

\subsection{Tadpole at finite temperature}
For instance, let us first reconsider the $\langle\phi\rangle$ tadpole in \eqref{phi tadpole},
\begin{align*}
	\langle \phi \rangle_\beta^{(1)} = \lim_{\epsilon\rightarrow 0^+}\frac{\lambda}{m^2}\, \frac{1}{\beta}\sum_k\frac{e^{+ik\epsilon}}{-ik+\mu}.
\end{align*}
%Here a couple changes from the continuum case were made. This time the external momentum dependence $p$ is shown explicitly, along with a Kronecker delta $\delta_{p,0}$ indicating that this expectation value is only nonzero for vanishing $p$. Strictly speaking the continuum case in momentum space also involves an overall Dirac delta function which was not shown.
The integral over $k$ has been turned into a sum, since the antiperiodic condition on the fermion fields restricts momentum to discrete values $$k=\frac{2\pi}{\beta}\left(n+\frac{1}{2}\right).$$
For the sake of evaluation it will be useful to turn this sum back into an integral by using a Dirac comb,
\begin{align}
\frac{1}{\beta}\sum_n 2\pi \delta\left(k- \frac{2\pi}{\beta}\left(n+\frac{1}{2}\right)\right)=\sum _n (-1)^n e^{-ik\beta n}.\label{Dirac comb}
\end{align}
The exponential factors from the Dirac comb will be sufficient to provide regularization for $n\neq 0$, so the limit $\epsilon \rightarrow 0^+$ may be taken in advance. The $n=0$ term is the same as the continuum case, and vanishes for $\mu>0$,
\begin{align}
	\langle \phi\rangle_\beta^{(1)} &= \frac{\lambda}{m^2}\sum _{n\neq 0} (-1)^n\,\int\frac{dk}{2\pi}\frac{e^{-ik\beta n}}{-ik+\mu}=\frac{\lambda}{m^2}\sum _{n> 0} (-e^{-\mu\beta})^n\non
	&=\left( -\frac{\lambda}{m^2}\right)\frac{e^{-\mu\beta}}{1+e^{-\mu\beta}}.\label{finite temperature tadpole}
\end{align}
This agrees with the quantum mechanical picture since this is just the shift in $\phi$ in the fermionic sector \eqref{qF} weighted by approximately the correct Boltzmann factor. Higher order corrections in $\lambda$ will just correct the bare mass $\mu$ in the Boltzmann factor to the exact value $\mu_\lambda$. Note also that the above calculation was for $\mu>0$, but this will also work out for the $\mu<0$ case if the non-vanishing $n=0$ term is taken into account.
\subsection{A puzzle involving the two-point function}\label{section puzzle}
Now let us return to the corrections of the bosonic two-point function in Fig. \ref{Fig2}, which was the main motivation for considering the situation at finite temperature. First let us consider the exact answer from the quantum mechanical perspective.

At $\lambda=0$ the finite temperature harmonic oscillator correlation function is given by (taking for simplicity $\beta>\tau>0$)
\begin{align}
	\langle q(\tau)q(0)\rangle^{(0)}_{\beta}=\frac{1}{2m}\frac{e^{-m\tau}+e^{-m\left(\beta-\tau\right)}}{1-e^{-m\beta}}.\label{finite temp correlation zeroth order}
\end{align}
This may be found by considering the trace over eigenstates $|n\rangle$ as in \eqref{thermal expectation}, and considering $q$ as a sum of harmonic oscillator ladder operators $q=\frac{1}{\sqrt{2m}}\left(a+a^
\dagger\right)$. It may also be found in the path integral approach by finding the Fourier transform of the propagator $(p^2+m^2)^{-1}$ with $p$ only taking discrete values $p=\frac{2\pi}{\beta}n$ as required by the periodic boundary conditions.

Note that similarly to the zero-temperature case \eqref{propagator phi}, the connected finite temperature two-point function is insensitive to an overall shift in the spectrum. Only the differences between eigenvalues matter. So considering the more general $\lambda\neq 0$ case, the shift in energy $\mu_\lambda$ in the fermionic sector will not matter at all. In both sectors the $\tau$ dependence is entirely given by the uncorrected $\lambda=0$ correlation function above.

The only appearance of $\lambda$ is through the disconnected correlation function as in diagram B of Fig. \ref{Fig2}. This is only nonvanishing in the fermionic sector, so after weighting it by the appropriate Boltzmann factor the exact two-point correlation function should be
\begin{align}
\langle q(\tau)q(0)\rangle_{\beta}=	\langle q(\tau)q(0)\rangle^{(0)}_{\beta}+\left(\frac{\lambda}{m^2}\right)^2\frac{e^{-\beta \mu_\lambda}}{1+e^{-\beta \mu_\lambda}}.\label{finite temp correlation}
\end{align}

But here we face some difficulties. Diagram A of Fig. \ref{Fig2} will be non-vanishing for finite temperature. But the only part of our correlation function which should depend on external momentum is the zeroth order piece $\langle q(\tau)q(0)\rangle^{(0)}_{\beta}$ which is already given fully by the bare propagator $(p^2+m^2)^{-1}$. Furthermore, Diagram B seems like it could give the correct $\lambda$ dependent disconnected term, but if we square the tadpole diagram in \eqref{finite temperature tadpole} the Boltzmann factor will also be squared, so it does not agree with the exact answer \eqref{finite temp correlation}.

Not surprisingly these two problems will end up being related to each other, and they may be clarified by generalizing the problem somewhat. The perturbation series in $\lambda$ may be expressed from the operator point of view in terms of the Dyson series in the interaction picture. General correlation functions of $\phi$ may be written as expectation values in the non-interacting system,
\begin{align*}
	\langle \phi(\tau_1)\phi(\tau_2)\dots\rangle_\beta = \frac{1}{Z}\sum_{n} e^{-E^{(0)}_n\beta}\langle n_0|\,\mathcal{T} e^{-\lambda\int_{0}^\beta d\tau \,q_0(\tau)N_0(\tau)}q_0(\tau_1)q_0(\tau_2)\dots\,|n_0\rangle.
\end{align*}
Here the states $|n_0\rangle$ are eigenstates of the free Hamiltonian $H_0$, in which the bosonic operator $q_0$ and the fermionic number operator $N_0$ (both with time dependence given by $H_0$) are completely decoupled. So upon expanding the exponential, the expectation value over the bosonic and fermionic fields may be factored, and the fermionic factor will involve products of number operators like $\langle \mathcal{T}\,N_0(\tau'_1)N_0(\tau'_2)N_0(\tau'_3)\dots \rangle_\beta$.

\begin{figure}
	\centering
	\includegraphics[width=0.8\textwidth]{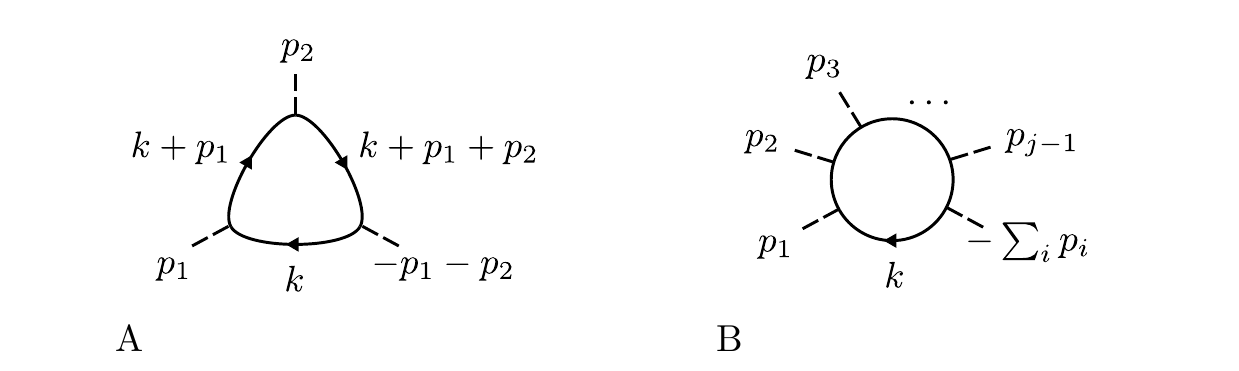}
	\caption{Higher order connected fermion loops. Diagram A involves three insertions of $\bar{\psi}\psi$. There is one additional inequivalent diagram in which external momentum $p_1$ and $p_2$ are interchanged. Diagram B involves $j$ insertions. There are are $(j-1)!$ distinct permutations of external momenta. }\label{Fig3}
\end{figure}

From the path integral point of view these correlation functions of number operators form fermion loops, both connected and disconnected, exactly as in Fig. \ref{Fig2} and more generally in Fig. \ref{Fig3}. But from the operator point of view, these correlation functions are extremely simple. $N$ commutes with the Hamiltonian so there is no time dependence, and from the algebra of fermionic ladder operators it is true that arbitrary powers $N^k$ reduce to $N$ itself. So no matter how many number operators at arbitrary times are inserted, the correlation function always equals the same factor,
\begin{align}
\langle N(\tau'_1)N(\tau'_2)N(\tau'_3)\dots \rangle^{(0)}_\beta = \frac{e^{-\mu\beta}}{1+e^{-\mu\beta}}.\label{finite temperature fermion loops}
\end{align}

How can this be possible from the path integral point of view, where this correlation function is expressed in terms of fermion loops which naively appear to be dependent on external momentum? Answering this question will be the focus of the next section.

\subsection{Resolution and fermion loops to all orders}\label{section fermion loops all-orders}

Let us first return to the correction to the two-point propagator $D(p)^{(2)}_\text{connected}$ in Fig. \ref{Fig2} and \eqref{D connected},
\begin{align*}
	D^{(2)}_\text{connected}(p)=\frac{\lambda^2}{\left(p^2+m^2\right)^2}\frac{1}{\beta}\sum_k \left(\frac{1}{k+i\mu}\right)\left(\frac{1}{k-p+i\mu}\right).
\end{align*}
The sum over discrete momenta may be transformed to a sum over continuous integrals of momenta, as was done for the $\phi$ tadpole above \eqref{Dirac comb}. We will assume here for simplicity that $\mu>0$,
\begin{align*}
	D^{(2)}_\text{connected}(p)&=\frac{\lambda^2}{\left(p^2+m^2\right)^2}\sum_{n>0} (-e^{-\mu\beta})^n\int \frac{dk}{2\pi} \frac{e^{-i\left(k+i\mu\right)\beta n}}{\left(k+i\mu\right)\left(k-p+i\mu\right)}\non
	&=\frac{\lambda^2}{\left(p^2+m^2\right)^2}\sum_{n>0} (-e^{-\mu\beta})^n\,\frac{i}{p}\left(1-e^{-ip\beta n}\right).
\end{align*}
Here we come to one of the problems discussed last section. The correction appears to be dependent on external momentum $p$, but from the quantum mechanics it is seen that there are only constant corrections to the two-point function.

But this actually has an easy resolution. If the finite temperature is treated consistently $p$ must itself by discretized in a periodic way $p=\frac{2\pi}{\beta}n$. So $1-e^{-ip\beta n}=0$, and the correction vanishes. This is not yet quite the whole resolution since the disconnected Diagram B in Fig. \ref{Fig2} does not equal the expected constant. However $D^{(2)}_\text{connected}(p=0)$ does not vanish since the $1/p$ factor makes the expression indeterminate. Formally taking the limit $p\rightarrow 0$ to calculate the double pole,
\begin{align}
	D^{(2)}_\text{connected}(p)&=\beta \delta_{p,0}\,\frac{-\lambda^2}{m^4}\sum_{n>0} n(-e^{-\mu\beta})^n=\beta \delta_{p,0}\,\frac{\lambda^2}{m^4}\frac{e^{-\mu\beta}}{\left(1+e^{-\mu\beta}\right)^2}.
\end{align}
The $\beta \delta_{p,0}$ factor is simply the overall delta function which a constant in $\tau$ picks up when Fourier transformed. So considering $D^{(2)}(\tau)$ instead, and adding the disconnected contribution which is the square of \eqref{finite temperature tadpole},
\begin{align}
	D^{(2)}(\tau)&=\frac{\lambda^2}{m^4}\frac{e^{-\mu\beta}}{\left(1+e^{-\mu\beta}\right)^2}+\left( -\frac{\lambda}{m^2}\frac{e^{-\mu\beta}}{1+e^{-\mu\beta}}\right)^2=\frac{\lambda^2}{m^4}\frac{e^{-\mu\beta}}{1+e^{-\mu\beta}},
\end{align}
and this is exactly what is needed from quantum mechanical considerations \eqref{finite temp correlation}. Higher order corrections in $\lambda$ will further correct $\mu$ to $\mu_\lambda$.

This resolves the issue for the bosonic two-point function, but it may be enlightening to consider higher order fermion loops, which by \eqref{finite temperature fermion loops} somehow must all evaluate to the same constant factor.

Consider first the three-point case as in Diagram A of Fig. \ref{Fig3}. Once again the discrete sum may be brought into the form of an integral, which now has an extra pole compared to the two-point case,
\begin{align*}
\int \frac{dk}{2\pi} \frac{e^{-i\left(k+i\mu\right)\beta n}}{\left(k+i\mu\right)\left(k+p_1+i\mu\right)\left(k+p_1+p_2+i\mu\right)}=-i\left(\frac{1}{p_1\left(p_1+p_2\right)}-\frac{e^{ip_1\beta n}}{p_1p_2}+\frac{e^{i(p_1+p_2)\beta n}}{p_2\left(p_1+p_2\right)}\right).
\end{align*}
Once again if $p_1, p_2, p_1+p_2$ are all non-zero, the exponential factors in this expression may be set to unity and the correction vanishes. However if one of these external momenta vanishes but not the others (say $p_1=0, p_2\neq 0$), then it can be shown taking the limit that the expression does not vanish and depends on external momentum $p_2$.

The resolution this time is that there are two distinct ways to pair the fermion fields in the three-point function. Besides the ordering of external legs shown in Fig. \ref{Fig3} there is a diagram with $p_1$ and $p_2$ interchanged, and the external momentum dependence of this diagram can be shown to fully cancel with the dependence of the expression above.

At higher order a similar cancelation of external momentum dependence holds. For instance, the evaluation of four-point loops leads to poles not only in the external momenta, but also in their sums such as $p_1+p_3$. Once again, if all sums of combinations of external momenta are non-zero, the individual diagram straightforwardly evaluates to zero. Otherwise, if some combinations vanish but not all, then there is a more complicated cancellation between the $3!$ distinct pairings of the fermion fields.

So in summary, for a general connected $j$-point fermion loop such as Diagram B of Fig. \ref{Fig3}, the only non-vanishing contribution will be when all external momenta $p_i=0$. Given this statement, it is possible to inductively prove \eqref{finite temperature fermion loops} which states that each $j$-point function of number operators evaluates to the same factor.

The full $j$-point function will involve a sum over disconnected diagrams, but first let us consider a connected $j$-point loop as in Diagram B. The evaluation is very similar to the two-point function at the beginning of this section. There is a factor of $i$ for each propagator, an overall minus sign for the fermion loop, a factor of $(j-1)!$ from summing over permutations of external momenta, and a factor of $\beta$ from the overall momentum conservation delta function,
\begin{align*}
	\left\langle N(p=0)^j\right\rangle_\text{connected}&=-\beta\, i^j(j-1)!\sum _{n>0} (-e^{-\mu\beta})^n\int \frac{dk}{2\pi} \frac{e^{-i\left(k+i\mu\right)\beta n}}{\left(k+i\mu\right)^j} \non
	&= -\beta^{j}\sum _{n>0} n^{j-1}(-e^{-\mu\beta})^n.
\end{align*}
As discussed for the two-point function, the factors $\beta^j$ arise from Fourier transforming a constant function of $j$ Euclidean time arguments, and will not appear in real space.

%\footnote{In the two-point case, one of the time arguments was held fixed at zero rather than being Fourier transformed. This provided an extra sum over Fourier components that canceled with the overall delta function.}

Now suppose inductively that for all $i<j$,
$$\left\langle N(\tau_1)N(\tau_2)\dots N(\tau_i) \right\rangle=\frac{e^{-\mu\beta}}{1+e^{-\mu\beta}}.$$
Then $\left\langle N(\tau_1)\dots N(\tau_j)\right\rangle$ may be split into a connected $j$-point loop diagram, and a series of diagrams where the first $N(\tau_1)$ operator is part of a connected $i$-point loop and the full expectation of the operators not on the loop is ${e^{-\mu\beta}}/{\left(1+e^{-\mu\beta}\right)}$ by the induction step.
\begin{align*}
	\left\langle N(\tau_1)\dots N(\tau_j)\right\rangle&=-\sum _{n>0} (-e^{-\mu\beta})^n\left[n^{j-1}+\frac{e^{-\mu\beta}}{1+e^{-\mu\beta}}\sum_{i=1}^{j-1}\left(\begin{array}{c}
		j-1\\ i-1
	\end{array}\right) n^{i-1}\right]\non
&=-\sum _{n>0} (-e^{-\mu\beta})^n\left[n^{j-1}+\frac{e^{-\mu\beta}}{1+e^{-\mu\beta}}\left(\left(n+1\right)^{j-1}-n^{j-1}\right)\right]\non
&=-\frac{1}{1+e^{-\mu\beta}}\sum_{n>0}\left[n^{j-1}(-e^{-\mu\beta})^n-\left(n+1\right)^{j-1}(-e^{-\mu\beta})^{n+1}\right].
\end{align*}
This is a telescoping series and only the $n=1$ term survives. So indeed for all $j$,
$$\left\langle N(\tau_1)\dots N(\tau_j) \right\rangle=\frac{e^{-\mu\beta}}{1+e^{-\mu\beta}},$$
which is in accordance with the behavior of fermionic number operators in ordinary quantum mechanics \eqref{finite temperature fermion loops}.
\section{Conclusion} \label{section conclusion}
Now let us briefly recap what has been shown here. We have just finished a calculation in Sec. \ref{section fermion loops all-orders} showing how fermion loops of arbitrary order are consistent with interpretation of $\bar{\psi}\psi$ as a number operator of a fermionic two-level system. This calculation was not specific to our model with a Yukawa interaction term. The same fermion loops appear in calculating correlation functions of number operators even in a system of free fermions, so this calculation is really a statement about how fermions  work in one spacetime dimension in general.

We began considering this calculation because of the puzzle raised in Sec. \ref{section puzzle}. It was shown from the operator perspective  in \eqref{finite temp correlation} that the Yukawa interaction should only correct the bosonic correlation function by a constant shift, even in the case of non-zero temperature. But before completing the calculation it was difficult to see how all the corrections to the bosonic correlation function involving fermion loops could possibly be independent of the $\tau$ parameter in the correlation function.

This puzzle involving fermion loop corrections to the bosonic correlation function was the main impetus for the considerations in this article. But as a pedagogical example which may supplement a course in quantum field theory, this need not be taken so far. Already in the introduction following Eq. \ref{Action}, it was shown how canonical quantization of the natural extension of the Yukawa interaction action to $d=1$ leads to a trivially solvable quantum mechanical system. In particular the two main effects of the interaction were shifting the energy gap of the two-level system from the bare value $\mu$ to the corrected value $\mu_\lambda$, and shifting the ground state expectation value of the harmonic oscillator position operator to some non-zero value when the two-level system is in its $N=1$ sector.

Both of these effects were shown from the path integral perspective in Sec.\ref{section Zero temperature} by calculating the two diagrams in Fig \ref{Fig1}. To find the shift in the ground state expectation value the tadpole diagram was calculated in Sec.\ref{section Tadpole}. To calculate the correction to the `mass' $\mu_\lambda$, the fermionic two-point function was calculated in Sec. \ref{section fermion 2 point}. It was hoped that presenting these calculations where the results have a clear quantum mechanical interpretation may improve understanding of similar calculations of vacuum expectation values and self-energy corrections in standard quantum field theory, without all the extra complications of renormalization.

\begin{acknowledgments}
	I would like to thank the Physics Stack Exchange community where I originally posed and answered these questions and was encouraged to publish them \cite{StackExchange}. I would like to thank the referees for their helpful comments improving the presentation of the paper.
\end{acknowledgments}

\end{spacing}

\end{document}